\documentstyle[preprint,aps]{revtex}
\tightenlines
\begin{document}
\draft
\title{ Decoherence and Relaxation of a Quantum Bit \\
in the Presence of Rabi Oscillations.}
\author{Anatoly Yu. Smirnov }
\address{ D-Wave Systems Inc. 320-1985 West Broadway,\\
 Vancouver, B.C. V6J 4Y3, Canada }
\date{\today}
\maketitle

\begin{abstract}
{Dissipative dynamics of a quantum bit driven by a strong resonant field and interacting with a heat
bath is investigated. We derive generalized Bloch equations and find modifications of the qubit's
damping rates caused by Rabi oscillations. Nonequilibrium decoherence and relaxation of a phase qubit inductively
coupled to a LC-circuit is considered as an illustration of the general results. It is 
argued that recent experimental results give a clear evidence of effective suppression of decoherence in a strongly driven flux qubit. }
\end{abstract}


\section{Introduction}

 Many years ago I. Rabi has shown \cite{Rabi1} that a population difference of a two-level
system (or a qubit in a modern language) subjected to the action of a resonant electromagnetic field
oscillates in time with a frequency proportional to the field strength. Since then the Rabi
oscillations and corresponding Rabi splitting of energetic levels have been observed in atomic
ensemble coupled to a single-mode cavity \cite{Raizen1}, in a single excitonic quantum dot
\cite{Stievater1,Kamada1,Htoon1,Zrenner1} as well as in a Cooper-pair box \cite{Nakamura1} and in a
Josephson junction \cite{Yu1}. A successful demonstration of Rabi oscillations in a Cooper-pair box
combined with Josephson junctions \cite{Vion1} along with completed and projected experiments on a
coherent manipulation of flux qubits \cite{Wal1,Crankshaw1,Chior2003,Greenberg1} offer strong possibilities for
a practical realization of a superconducting quantum bit\cite{Makhlin1}. This objective can be
accomplished by suppressing noise and decoherence caused by a dissipative environment (a heat
bath).

Corresponding decoherence and relaxation times, $T_2$ and $T_1,$ are usually expressed in terms
of the spectral density of the heat bath fluctuations, $J(\omega),$ which represents yet an imaginary
part of a heat bath susceptibility $\chi(\omega), J(\omega) = \chi''(\omega)$
\cite{Slichter1,Grifoni1,Tian1}
\begin{eqnarray}
T_1^{-1} = {\Delta^2\over 2\omega_0^2}J(\omega_0)\coth\left({\omega_0 \over 2T}\right), \nonumber\\
T_2^{-1} = {1\over 2} T_1^{-1} + {\epsilon^2 \over 2\omega_0^2} \left[J(\omega)\coth\left({\omega
\over 2T}\right)\right]_{\omega = 0}.
\end{eqnarray}
Here $\Delta$ and $\epsilon$ are the tunnel splitting and the energy bias of the qubit, $\omega_0 =
\sqrt{\Delta^2 + \epsilon^2},$ and $T$ is temperature of the environment. Henceforward the Boltzmann
and Planck constants are supposed to be unity, $\hbar =1, k_B = 1,$ unless otherwise specified. A
combination $S(\omega) = J(\omega)\coth(\omega / 2T)$ stands here for the spectrum of heat bath
fluctuations. Spectra of additional noise sources should be properly added to $S(\omega)$ in Eq.(1),
in particular, a spectrum of $1/f$ (flicker) noise, $S_f(\omega) \sim 1/|\omega|,$ which is inherent to
every electrical circuit. These circuits are necessary to control and read out the state of the
superconducting qubit. It follows from Eq.(1) that in the presence of energy bias $(\epsilon\neq 0)$
$1/f$-noise gives an infinite contribution to the transversal rate $T_2^{-1}$ that leads to very fast
dephasing of the qubit. Obviously a formal application of Eqs.(1) to the case of $1/f$ noise can not
be justified because the formulas (1) are valid only for the weak interaction between the qubit and
the heat bath. An additional point to emphasize that the above-mentioned expressions present the rates
of qubit's relaxation to the thermodynamically-equilibrium state in the absence of any time-dependent
external force acting on the qubit. With the driving force the relaxation and decoherence rates are
modified as it is demonstrated by F. Bloch \cite{Bloch1} for a spin $1/2$ in a rotating magnetic
field. This phenomenon opens the road for the suppression of decoherence by the external
time-dependent field which is especially effective when the specific time scales describing the
internal qubit's dynamics are less than the correlation time $\tau_c$ of the heat bath. The similar
mechanism underlies a concept of a quantum "bang-bang" control, proposed by Viola {\it et al}
\cite{Viola1,Viola2}. It was shown that a sequence of microwave pulses can completely suppress an
environment-induced decoherence of the qubit if the repetition frequency of the pulses exceeds the
inverse correlation time of the heat bath. The general concept is exemplified by an exactly solvable
model of a two-level system with energy splitting fluctuating due to interaction with a heat bath.

Effects of the strong driving force on decoherence and relaxation of the dissipative two-level system were under extensive discussion for many years 
\cite{Grifoni1,Grifoni2,Weiss1,Shao1}. Bloch-Redfield type equations with an arbitrary control field have been derived \cite{Hartmann1} in the process. However, in addition to the general theory there is an urgent need now in the transparent analytical formulas for the decoherence and relaxation rates of the resonant-driven quantum bit.  

In the present paper we analyze effects of Rabi oscillations on decoherence and relaxation of a two-level system (a qubit) with the goal to find field-induced modifications of the formulas (1). To do that, we resort to the theory of open quantum systems \cite{Efremov1,Smirnov1,Rose1} as well as to the rotating wave approximation. This approximation is quite justified here, because for the most experimental situations the frequency of the Rabi oscillations induced by the resonant driving force is much less than the energy splitting of the qubit. 

The paper is
organized as follows. In Sec.II we derive non-Markovian equations for the qubit's variables in the
rotating frame and bring them to the form of the generalized Bloch equations with kinetic coefficients
depending on the Rabi frequency. Dissipative dynamics of the qubit is described in Sec.III. We
calculate the nonequilibrium damping rates and illustrate the results using the flux qubit coupled to
a LC-circuit (a tank) as an example. A theoretical disccussion of recent experimental results \cite{Chior2003} for dephasing and relaxation times of strongly driven flux qubit is presented in Sec. IV. 

\section{ Generalized Bloch equations}

The Hamiltonian of the qubit interacting with a resonant field $F(t)$ and coupled to a dissipative
environment with a variable $Q(t)$ has the form
\begin{equation}
H = {\Delta\over 2} \sigma_x + {\epsilon \over 2} \sigma_z - \sigma_z F \cos\omega_0t - {1\over 2}
\sigma_z Q,
\end{equation}
where $\Delta $ is a tunneling rate of the qubit, $\epsilon$ is a bias, and $F\cos\omega_0t$ is a
resonant driving force with the frequency exactly equal to the energy splitting of the qubit:
$\omega_0 = \sqrt{\Delta^2 + \epsilon^2}$. Here we have some distinctions from the Hamiltonian
describing a dissipative evolution of a spin in a rotating magnetic field \cite{Bloch1,Shao1,Smirnov1}.
First of all, in the present formulation of the problem the alternating  external magnetic field is
aligned with one axis $z.$ Because of this equations describing a free evolution of the qubit (without
its coupling to the heat bath) do not have exact solutions and we have to resort to the rotating wave
approximation. Besides that we have an energy bias $\epsilon$ in Eq.(2), and the heat bath variables
$Q$ are coupled only to z-direction of the qubit's Pauli matrix. For the superconducting flux qubit
\cite{Mooij1,Orlando1,Crankshaw1} the parameters $\epsilon, F$ and $Q$ can be considered as static,
oscillating and fluctuating parts of the flux produced by the external circuit.

It is convenient to introduce the following operators describing the qubit in the rotating frame of
reference:
\begin{eqnarray}
X = {\Delta \over \omega_0}\sigma_x + {\epsilon \over \omega_0}\sigma_z, \nonumber\\
Y = \sigma_y \cos\omega_0 t + \left({\Delta \over \omega_0}\sigma_z - {\epsilon \over
\omega_0}\sigma_x\right)\sin\omega_0 t, \nonumber\\
Z =  \left({\Delta \over \omega_0}\sigma_z - {\epsilon \over \omega_0}\sigma_x\right) \cos\omega_0 t -
\sigma_y \sin\omega_0 t.
\end{eqnarray}
These operators  explicitly depend on time, but preserve all commutation rules inherent to Pauli
matrices,for example, $[X,Y]_- = 2iZ, $ and so on. In the rotating wave approximation (RWA) the effective Rabi frequency of the qubit $\Omega_R = (\Delta/\omega_0)F$ is much less than the frequency of the driving force $\omega_0, \Omega_R \ll \omega_0,$ so that a parameter $\alpha, \alpha = \Omega_R /4\omega_0,$ is much less than one, $\alpha \ll 1.$ Then, the Hamiltonian (2) of the system has the form:
\begin{equation}
H = {\omega_0 + \alpha \Omega_R\over 2} X - {\Omega_R \over 2} Z - {1\over 2} \left\{{\Delta \over \omega_0} \left(Z
\cos \omega_0t + Y \sin \omega_0t\right) + {\epsilon \over \omega_0} X\right\}Q.
\end{equation}
Here we take into account first-order corrections to the RWA described by the parameter $\alpha$ and resulting in the frequency shift of the two-state system.  
In the absence of the driving force and dissipation the Hamiltonian of the qubit is proportional to the operator $X$. 
Then, the time evolution of diagonal elements of the qubit's density matrix in the energy representation is determined by $X$, whereas the operators $Y$ and $Z$ determine the evolution of the off-diagonal elements of the density matrix. With the driving force the situation is essentially complicated. Nevertheless, by convention the decay time of the operator $X$ will be subsequently referred to as a relaxation time, and the decay time of the operator $Z$ will be referred to as a decoherence time \cite{Nielsen1}.

We resort to the rotating wave approximation  and drop fast
oscillating terms. Then the time evolution of the Heisenberg operators $X,Y,Z $ is governed by the
following equations:
\begin{eqnarray}
\dot{X} = \Omega_R Y + {\Delta \over \omega_0}\left(Y \cos \omega_0t - Z \sin \omega_0t\right)Q,
\nonumber\\
\dot{Y} = -\Omega_R X -  \alpha \Omega_R Z - \left({\Delta \over \omega_0}X \cos \omega_0t -  {\epsilon \over
\omega_0}Z\right)Q, \nonumber\\
\dot{Z} = \alpha \Omega_R Y + \left({\Delta \over \omega_0}X \sin\omega_0t -  {\epsilon \over \omega_0}Y\right)Q.
\end{eqnarray}
In the absence of the qubit-bath interaction the variables $ X,Y$ oscillate in time,
\begin{eqnarray}
X(t) = X(t_1) \cos\Omega_R(t-t_1) +  Y(t_1) \sin\Omega_R(t-t_1) -\alpha Z(t_1)[1 - \cos \Omega_R(t-t_1)], \nonumber\\
Y(t) = Y(t_1) \cos\Omega_R(t-t_1) -  X(t_1) \sin\Omega_R(t-t_1) - \alpha Z(t_1) \sin \Omega_R(t-t_1), \nonumber\\
Z(t) = Z(t_1) + \alpha Y(t_1)\sin\Omega_R(t-t_1) - \alpha X(t_1)[1 - \cos \Omega_R(t-t_1)].  
\end{eqnarray}
Because of this, the parameter $\Omega_R,$
\begin{equation}
\Omega_R = {\Delta \over \sqrt{\Delta^2 + \epsilon^2}}F,
\end{equation}
represents the Rabi frequency. The dissipative terms in the equations (5) contain fast oscillating
components which should be subsequently omitted. In the case of the Gaussian fluctuations of the
unperturbed heat bath variables $Q^{(0)}$ or for weak coupling between the qubit and the heat bath the
total operator $Q(t)$ is a linear functional of the qubit's variables\cite{Efremov1,Puller1}:
\begin{equation}
Q(t) = Q^{(0)}(t) + {1\over 2}\int dt_1 \varphi (t,t_1)\left\{{\Delta \over \omega_0} \left(
Z(t_1)\cos\omega_0t_1 + Y(t_1)\sin \omega_0t_1\right) + {\epsilon \over \omega_0} X(t_1)\right\}.
\end{equation}
Here
\begin{equation}
\varphi (t,t_1) = \langle i [Q^{(0)}(t),Q^{(0)}(t_1)]_-\rangle \theta (t-t_1)
\end{equation}
is a linear response function of the heat bath (a retarded Green function). By averaging the equations
over the initial state of the heat bath with the temperature $T$ we obtain the system of the
non-Markovian equations:
\begin{eqnarray}
\langle \dot{X} \rangle = \Omega_R \langle Y \rangle  + {\Delta \over 2 \omega_0^2} \int dt_1 \{
\tilde{M}(t,t_1)\langle i[Y(t)\cos\omega_0t - Z(t) \sin\omega_0t, \nonumber\\
\Delta Z(t_1)
\cos\omega_0t_1 +
\Delta Y(t_1)\sin\omega_0t_1 + \epsilon X(t_1)]_-\rangle + \nonumber\\
\varphi(t,t_1)\langle (1/2)[Y(t)\cos\omega_0t - Z(t) \sin\omega_0t, \nonumber\\
\Delta Z(t_1)
\cos\omega_0t_1 + \Delta Y(t_1)\sin\omega_0t_1 + \epsilon X(t_1)]_+\rangle \},
\end{eqnarray}
\begin{eqnarray}
\langle \dot{Y} \rangle = -\Omega_R \langle X \rangle  -  \alpha \Omega_R \langle Z \rangle  - {1 \over 2 \omega_0^2} \int dt_1 \{
\tilde{M}(t,t_1)\langle i[\Delta X(t)\cos\omega_0t - \epsilon Z(t), \nonumber\\
\Delta Z(t_1)
\cos\omega_0t_1 +
\Delta Y(t_1)\sin\omega_0t_1 + \epsilon X(t_1)]_-\rangle + \nonumber\\
\varphi(t,t_1)\langle (1/2)[\Delta X(t)\cos\omega_0t - \epsilon Z(t), \nonumber\\
\Delta Z(t_1)
\cos\omega_0t_1 + \Delta Y(t_1)\sin\omega_0t_1 + \epsilon X(t_1)]_+\rangle \},
\end{eqnarray}
\begin{eqnarray}
\langle \dot{Z} \rangle =  \alpha \Omega_R \langle Y\rangle + {1 \over 2 \omega_0^2} \int dt_1 \{ \tilde{M}(t,t_1)\langle i[\Delta
X(t)\sin\omega_0t - \epsilon Y(t), \nonumber\\
\Delta Z(t_1) \cos\omega_0t_1 +
\Delta Y(t_1)\sin\omega_0t_1 + \epsilon X(t_1)]_-\rangle + \nonumber\\
\varphi(t,t_1)\langle (1/2)[\Delta X(t)\sin\omega_0t - \epsilon Y(t), \nonumber\\
\Delta Z(t_1)
\cos\omega_0t_1 + \Delta Y(t_1)\sin\omega_0t_1 + \epsilon X(t_1)]_+\rangle \}.
\end{eqnarray}
Here $\tilde{M}(t,t_1) = M(t,t_1)\theta(t-t_1),$ and
\begin{equation}
M(t,t_1) =\langle (1/2) [Q^{(0)}(t),Q^{(0)}(t_1)]_+\rangle
\end{equation}
is the symmetrized correlation function of the heat bath fluctuations, $\hbar=1, k_B=1.$ Henceforward
we use notations $[...]_- $ and $[...]_+$ for commutators and anticommutators, respectively, and
$\theta(\tau)$ is the Heaviside step function. It should be noted that the spectral density of heat
bath fluctuations,
\begin{equation}
S(\omega) = \int d\tau e^{i\omega \tau } M(\tau),
\end{equation}
is related to the imaginary part $\chi^{\prime\prime}(\omega ) = Im \chi(\omega ) = J(\omega )$ of the
susceptibility
\begin{equation}
\chi(\omega ) = \int d\tau e^{i\omega \tau }\varphi(\tau )
\end{equation}
according to the fluctuation-dissipation theorem
\begin{equation}
S(\omega ) = \chi^{\prime\prime}(\omega ) \coth \left({\omega \over 2T}\right).
\end{equation}
To derive the system (10),(11),(12)  we resort to the general theory of open quantum systems developed
in Refs. \cite{Efremov1,Smirnov1,Rose1,Puller1}. These equations are valid exactly in the case of
Gaussian fluctuations of the unperturbed variables of the heat bath $ \{Q^{(0)}(t)\}$ or approximately for the weak coupling between the qubit and the heat bath. 
In the present paper we restrict ourselves to the Bloch-Redfield approximation when the qubit-bath interaction is not strong, so that relaxation of the qubit during the correlation time $\tau_c$ of the
heat bath is negligible, whereas the period of Rabi oscillations, $2\pi /\Omega_R,$ can be of order of $\tau_c: \Omega_R \tau_c \sim 1.$ Generally speaking, the perturbation theory is not quite applicable to the case of 1/f spectrum of heat bath noise as well as to the case of resonant coupling of the qubit to the high quality LC circuit having a peaked spectral density. But even for these cases the perturbative approach allows us to find conditions when the contribution of the environment to the dissipative dynamics of the qubit will be moderate, thus justifying using the Bloch-Redfield approximation. 

The non-Markovian equations (10)-(12) contain the commutators of the qubit's operators taken at different moments of time $t$ and $t_1$ separated by the correlation time of the heat bath $\tau_c$. To calculate the commutators we express the operators $X(t_1), Y(t_1),$ and $Z(t_1)$ in terms of the operators $X(t), Y(t),$ and $Z(t)$ using the  
free evolution of the "dressed" qubit (6). It should be noted that reducing  the operators $X(t), Y(t),$ and $Z(t)$ to the operators $X(t_1), Y(t_1), Z(t_1)$ results in the same results for the qubit's damping rates $\Gamma_x,\Gamma_y, \Gamma_z.$ The operators of the qubit $X,Y,Z$ taken at the same moment of time $t$  obey the well-known commutation rules for the Pauli matrices, $i[X(t),Y(t)]_- = -2Z(t)$, and so on. 
Then, the commutators, like $i[X(t),Y(t_1)]_-$, oscillate with the Rabi frequency $\Omega_R$ as functions of the time difference $\tau = t - t_1 :$
$$ i[X(t),Y(t_1)]_- = - 2 Z(t) \cos \Omega_R \tau + 2 \alpha Y(t) \sin\Omega_R\tau.$$ 
The same is true for the anticommutators: $ (1/2)[X(t),Y(t_1)]_+ = \sin\Omega_R\tau.$   
These (anti)commutators should be substituted into Eqs.(10)-(12). Integrating over $\tau $ gives us the Fourier transforms of the correlation function (14) and the response function (15) of the dissipative environment taken at the frequencies $\omega_0, \omega_0\pm \Omega_R, $ and $\Omega_R.$ 
After averaging the collision integrals in Eqs.(10)-(12) over the fast oscillations with the frequencies $\omega_0, 2\omega_0, ...$ ($\cos\omega_0t \cos\omega_0t_1 \simeq (1/2)\cos\omega_0\tau,$ ...) and omitting unimportant terms we obtain the
generalized Bloch equations for the qubit's variables averaged over the initial state of the heat
bath:
\begin{eqnarray}
\dot{X} + \Gamma_x X = \Omega_R Y - \nu_x, \nonumber\\
\dot{Y} + \Gamma_y Y = -\Omega_R X -\alpha \Omega_R Z  + \nu_y, \nonumber\\
\dot{Z} + \Gamma_z Z = \alpha \Omega_R Y + \nu_z.
\end{eqnarray}
Here we drop the brackets $\langle ..\rangle $ symbolizing averaging over the heat bath fluctuations.
It should be mention also that in order to find the physical values we have to average the operators
$X,Y,Z$ over the initial state of the qubit. The relaxation coefficients in Eqs.(17) are determined by
the spectrum of heat bath fluctuations $S(\omega)$ (14) taken at the frequencies $\omega_0,
\omega_0\pm \Omega_R, \Omega_R:$
\begin{eqnarray}
\Gamma_x = {\Delta^2 \over 4 \omega_0^2} \left[ S(\omega_0) + {(1+2\alpha )S(\omega_0 +\Omega_R) + (1-2\alpha )S(\omega_0 -
\Omega_R) \over 2}\right], \nonumber\\
\Gamma_y = {\Delta^2\over 4\omega_0^2}\left[ S(\omega_0) + \alpha  {S(\omega_0 +\Omega_R) - S(\omega_0 - \Omega_R) \over 2}\right]  +  {\epsilon^2\over 2\omega_0^2}
S(\Omega_R), 
\end{eqnarray}
\begin{eqnarray}
\Gamma_z = {\Delta^2\over 4\omega_0^2}{(1+\alpha )S(\omega_0 +\Omega_R) + (1-\alpha )S(\omega_0 - \Omega_R)
\over 2} + {\epsilon^2\over 2\omega_0^2} S(\Omega_R).
\end{eqnarray}
The parameters $\nu_x,\nu_y,\nu_z$  defining the steady-state values of the qubit's operators  are
proportional to the imaginary and real parts of the heat bath susceptibility $\chi(\omega)$ (15):
\begin{eqnarray}
\nu_x = {\Delta^2 \over 4 \omega_0^2} \left[ \chi^{\prime\prime}(\omega_0) +
{(1+2\alpha )\chi^{\prime\prime}(\omega_0 +\Omega_R) + (1-2\alpha ) \chi^{\prime\prime}(\omega_0 -
\Omega_R) \over 2}\right], \nonumber\\
\nu_y = {\Delta^2 \over 4 \omega_0^2}\left[ {(1-\alpha )\chi^{\prime}(\omega_0 -\Omega_R) - (1+\alpha) \chi^{\prime}(\omega_0 + \Omega_R) \over 2} + \alpha \chi^{\prime}(\omega_0)\right] - \alpha  {\epsilon^2\over 2\omega_0^2} \left[ \chi^{\prime}(0) - \chi^{\prime}(\Omega_R)\right], \nonumber\\
\nu_z =  {\Delta^2 \over 4\omega_0^2}\left[{(1+\alpha )\chi^{\prime\prime}(\omega_0 +\Omega_R) - (1-\alpha )\chi^{\prime\prime}(\omega_0 - \Omega_R) \over 2} -\alpha \chi^{\prime\prime}(\omega_0)\right] + {\epsilon^2\over 2\omega_0^2} \chi^{\prime\prime}(\Omega_R).
\end{eqnarray}
The above-mentioned expressions for the kinetic coefficients of the qubit have been derived with taking into account first corrections to the rotating wave approximation. 

\section{ Nonequilibrium dissipative dynamics of the qubit }

For the steady-state values of the qubit's operators $X_0,Y_0,Z_0$ in the rotating frame of reference
we find the following expressions
\begin{eqnarray}
X_0 = {\Omega_R \nu_y - \Gamma_y \nu_x \over \Omega_R^2 + \Gamma_x \Gamma_y}, \nonumber\\
Y_0 = {\Omega_R \nu_x + \Gamma_x \nu_y \over \Omega_R^2 + \Gamma_x \Gamma_y},
\end{eqnarray}
\begin{equation}
Z_0 =   { \Delta^2 [\chi^{\prime\prime}(\omega_0 + \Omega_R) - \chi^{\prime\prime}(\omega_0 -
\Omega_R)] + 4 \epsilon^2 \chi^{\prime\prime}(\Omega_R) \over \Delta^2 \left[S(\omega_0 + \Omega_R) +
S(\omega_0 - \Omega_R)\right] + 4 \epsilon^2 S(\Omega_R)}.
\end{equation}
The small corrections to the rotating-wave approximation have been omitted here.
In the equilibrium case, when $\Omega_R = 0, Y_0=0,Z_0=0,$ we obtain the well-known expression for the
difference of population $X_0, X_0 = -\nu_x/\Gamma_x = - \tanh(\omega_0/2T).$ At the same time, in the
strong nonequilibrium case with $\Omega_R \gg \Gamma $ the steady-state values of $X$ and $Y$ go to
zero: $X_0=0,Y_0=0,$ so that the populations of the excited and ground states of the qubit are
equalized under the action of the resonant field; in so doing the qubit is polarized along Z-axis. The
dissipative evolution of the operators $X,Y,Z$ from their initial values $X(0),Y(0),Z(0)$ is described
by the equations
\begin{eqnarray}
X(t) = X(0) e^{-t/T_1} \left(\cos\Omega_Ct - {\Gamma_x-\Gamma_y \over
2\Omega_C}\sin\Omega_Ct\right) +
\nonumber\\
X_0\left[ 1 - e^{-t/T_1}\right] \left(\cos\Omega_Ct - {\Gamma_x-\Gamma_y \over
2\Omega_C}\sin\Omega_Ct\right)
+ \nonumber\\
\left[Y(0) - Y_0\right] e^{-t/T_1} {\Omega_R \over \Omega_C}\sin\Omega_Ct, \nonumber\\
Y(t) = Y(0) e^{-t/T_1} \left(\cos\Omega_Ct - {\Gamma_x-\Gamma_y \over 2\Omega_C}\sin\Omega_Ct\right) +
\nonumber\\
X_0\left[ 1 - e^{-t/T_1}\right] \left(\cos\Omega_Ct - {\Gamma_x-\Gamma_y \over
2\Omega_C}\sin\Omega_Ct\right)
+ \nonumber\\
\left[X(0) - X_0\right] e^{-t/T_1} {\Omega_R \over \Omega_C}\sin\Omega_Ct,\nonumber\\
Z(t) = Z(0) e^{-\Gamma_z t} + Z_0\left( 1 - e^{-\Gamma_z t} \right).
\end{eqnarray}
Here
\begin{equation}
\Omega_C = \sqrt{ \Omega_R^2 - {1\over 4} (\Gamma_x - \Gamma_y)^2}
\end{equation}
is the effective Rabi frequency shifted because of relaxation processes, and $T_1^{-1},$
\begin{equation} T_1^{-1} = {\Gamma_x + \Gamma_y \over 2} = {\Delta^2 \over 4 \omega_0^2}  S(\omega_0)
+ {\Delta^2 \over 8 \omega_0^2}  {(1+3\alpha )S(\omega_0 +\Omega_R) + (1-3\alpha )S(\omega_0 - \Omega_R) \over 2} +
{\epsilon^2\over 4 \omega_0^2} S(\Omega_R),
\end{equation}
is the longitudinal damping rate (with non-RWA amendments) which determines the relaxation of the qubit energy. 
It is worth to mention that the fast oscillating terms omitted in the process of our  calculation, which is based on the rotating wave approximation, result in the corrections to the relaxation rate $T_1^{-1}$ (25) of order of the small parameter 
$3\alpha = (3\Omega_R /4\omega_0) \ll 1.$  

Under the strong driving conditions, $\Omega_R \gg \Gamma,$ the low-frequency noise contributes to the relaxation rate $T_1^{-1}$ as well.  
In the absence
of the driving force and the Rabi oscillations, when $F=0, \Omega_R = 0,$ the effective Rabi frequency
(24)  becomes imaginary, $\Omega_C = (i/2)(\Gamma_x - \Gamma_y),$ and damping of the qubit's
variables $X,Y,Z$ is determined by the corresponding  rates $\Gamma_x,\Gamma_y, \Gamma_z,$
respectively. In this equilibrium case the formulas (1) come into play; in so doing the parameter
$\Gamma_x$ is related to the longitudinal relaxation rate, $\Gamma_x= T_1^{-1},$ whereas $\Gamma_y $
and $\Gamma_z$ describes the decoherence of the qubit caused by its interaction with the heat bath:
$\Gamma_y = \Gamma_z = T_2^{-1}.$
Interestingly that with the driving force the dissipative evolution of the operators $X$ and $Z$ is also determined by the different time scales $T_1$ (25) and 
$\Gamma_z^{-1}$ (19).

Let's assume that in the initial moment of time, $t=0,$ the qubit was in the ground state of the
Hamiltonian $\omega_0 X/2$. Averaging over this state gives: $\langle X(0)\rangle = -1, \langle
Y(0)\rangle = 0, \langle Z(0)\rangle = 0.$ Then, with the strong nonequilibrium conditions, $\Omega_R
\gg \Gamma,$ the probability to find the qubit in the excited state, $P_E = (1 + \langle X\rangle
)/2$, demonstrates damped Rabi oscillations with the relaxation time $T_1$ (25)
\begin{equation}
P_E(t) = {1\over 2} \left( 1 - e^{-t/T_1} \cos\Omega_R t \right).
\end{equation}
At large times, $t \gg T_1,$ the excitation probability  $P_E$ tends to $1/2$ that points to the
equalization of level populations.

 The distribution of the qubit between left and right wells can be described by
the difference of probabilities to find the system in the left well, $P_L$, and in the right well,
$P_R: P_L - P_R = \langle \sigma_z\rangle, $ with the total probability $P_L + P_R =1.$  Then, for the
qubit, initially prepared in the ground state, this difference also oscillates in time, but with
several frequencies: $\omega_0, \omega_0\pm\Omega_R,$ and $ \Omega_R:$
\begin{eqnarray}
(P_L - P_R)(t) = \langle \sigma_z\rangle = {\Delta \over \omega_0} e^{-t/T_1} \sin\Omega_Rt
\sin\omega_0t + \nonumber\\
{\Delta \over \omega_0}Z_0 \left( 1 - e^{-\Gamma_z t} \right)\cos\omega_0t
- {\epsilon\over \omega_0 } e^{-t/T_1} \cos\Omega_R t.
\end{eqnarray}
In the initial moment of time we have: $P_L(0)- P_R(0) = - \epsilon/\omega_0 < 0,$ that means that the
qubit was shifted towards the right well because of the bias $\epsilon$. Without the bias the
difference between the probabilities to find the qubit in the left and right wells shows quantum
beatings with the frequency $\omega_0$ corresponding to the energy splitting, the envelope amplitude
therewith oscillates with the Rabi frequency $\Omega_R$:
\begin{equation}
P_L - P_R = e^{-t/T_1} \sin\Omega_Rt \sin\omega_0t + Z_0 \left( 1 - e^{-\Gamma_z t}
\right)\cos\omega_0t.
\end{equation}
It is of interest that the Rabi oscillations of the envelope are damped out for the time scale $T_1$
(25) of energy relaxation, whereas the steady-state polarization of the qubit along the direction of
the external oscillating field is established for the different time scale $\Gamma_z^{-1}$ (19). In the presence
of the bias the nonequilibrium damping rates $T_1^{-1}$ and $\Gamma_z$ are related as
\begin{equation}
T_1^{-1} = {1\over 2} \Gamma_z + {\Delta^2 \over 4 \omega_0^2} S(\omega_0).
\end{equation}
For the three-junctions qubit \cite{Wal1,Mooij1,Orlando1} a persistent current in the qubit loop is
proportional to $\sigma_z-$matrix: $ \hat{I} = I_q \sigma_z,$ so that the function
$(P_L-P_R)(t)=\langle \sigma_z\rangle$ describes flipping of the current and the flux in the qubit's loop. 

 It follows from Eqs.(19),(25) the Rabi oscillations have a profound impact on decoherence  caused by $1/f$ (flicker) noise. Contrary to the equilibrium case, in the presence of Rabi oscillations the dominant contribution of $1/f$ noise
 into damping rates of the biased quantum bit is {\it finite}:
 $(\Gamma_z)_f = 2(T_1^{-1})_f =(\epsilon^2/2 \omega_0^2)(\upsilon/\Omega_R),$ where $S_f(\omega) =
\upsilon /|\omega|$ is a spectrum of $1/f$ noise with a frequency-independent constant $\upsilon$. No wonder that we do not have any divergent contributions now, because the low-frequency flicker noise is averaged over the rapid Rabi oscillations of the qubit. 
Our perturbative approach is valid if the flicker noise constant $\upsilon $ is smal enough, so that the contribution of $1/f$ noise to the damping rate, $(\Gamma_z)_f$,  is much less than the Rabi frequency $\Omega_R$:
$(\epsilon^2/2 \omega_0^2)(\upsilon/\Omega_R^2) \ll 1.$   

Besides the clear physical meaning the significance of the expressions (19),(25),(26)-(29) lies in the fact that these formulas give us a simple description of the dissipative evolution of the various Rabi-oscillating quantum bits.  
It is important, in particular, for the experiments with the phase qubits \cite{Wal1,Mooij1,Orlando1,Greenberg1} coupled to a LC-circuit (tank). The resonant tank circuit serves as a control element but also as a source of noise with a frequency-dependent spectral function $S(\omega)$ (16). 
 A persistent current in the qubit
loop with an inductance $L_q$ can take two values $\pm I_q.$  The circuit has an inductance $L_T,$ a
capacitance $C_T$ and is characterized by the resonant frequency $\omega_T = 1/\sqrt{L_TC_T}$ and the
quality factor  $Q_T = \omega_T/\gamma.$ Here $\gamma = 1/(R_TC_T)$ is a linewidth of the circuit
having a resistance $R_T.$ The qubit and the tank are coupled by a mutual inductance $M_c = k\sqrt{L_T
L_q}$ with $k^2Q_T \simeq 0.1-1,$ so that an interaction term has the form: $V_{int} = - M_cI_pI_T
\sigma_z$ where $I_T$ is a current in the tank. It means that the heat bath operator $Q(t)$ involved
in the Hamiltonian (2) is proportional to the current $I_T: Q = 2 M_cI_qI_T.$ After calculating the
susceptibility and the spectrum of the LC-circuit we find the function $\chi^{\prime\prime}(\omega )$
(15):
\begin{equation}
\chi^{\prime\prime}(\omega ) =   4 k^2\omega_T^2L_q I_q^2 { \gamma \omega  \over (\omega^2 -
\omega_T^2 )^2 + \gamma^2 \omega^2 },
\end{equation}
The resonant character of the heat bath spectrum $S(\omega), S(\omega ) = \chi^{\prime\prime}(\omega ) \coth(\omega/2T),$  results in the dependence of the relaxation and decoherence rates of the qubit, $T_1^{-1}$ (25) and $\Gamma_z$ (19), on the Rabi frequency $\Omega_R$ (7). This dependence is especially pronounced at low temperatures, $T < \Omega_R.$ In particular, in the case when the tunneling
splitting $\omega_0 = \sqrt{\Delta^2 + \epsilon^2}$ is much more than the resonant frequency of the
tank, $\omega_0 \gg \omega_T, $ the energy distribution $P_E$ (26) and the distribution of the qubit
between wells $P_L-P_R$ (27) relaxes with the rate
\begin{eqnarray}
T_1^{-1} =  k^2\omega_T^2{L_q I_q^2\over \hbar}  {\gamma\Delta^2\over 4\omega_0^5}\left[ 4\coth\left({\omega_0
\over 2T}\right) + \coth\left({\omega_0-\Omega_R \over 2T}\right) + \coth\left({\omega_0+\Omega_R
\over
2T}\right)\right] + \nonumber\\
 k^2\omega_T^2L_q I_q^2  {\epsilon^2\over \omega_0^2}{ \gamma \Omega_R  \over (\Omega_R^2 -
\omega_T^2 )^2 + \gamma^2 \Omega_R^2 }\coth\left({\Omega_R \over 2T}\right).
\end{eqnarray}
It is known \cite{Wal1,Mooij1,Orlando1} that for the persistent current qubit the non-zero bias
$\epsilon$ is necessary for the measurements. At the same time the non-zero bias means the significant
contribution of the last term in Eq.(31) into the relaxation rate $T_1^{-1}$. This contribution can be
especially dangerous if the Rabi frequency $\Omega_R$ approaches the resonant frequency of the tank
$\omega_T: \Omega_R \simeq \omega_T.$ To estimate the relaxation time $T_1$ we consider the phase
qubit with
 the inductance $L_q = 10 pH,$ and the current $I_q = 400 nA$. Then, the flux
 created by the qubit's loop is evaluated as: $L_qI_q = 2\times 10^{-3} \Phi_0$, where $\Phi_0=h/2e$
 is the flux quantum, and $L_qI_q^2 = 1.6\times 10^{-24} J.$ In the case of the exact resonance
 between the Rabi frequency and the frequency of the tank, $\Omega_R = \omega_T,$ the major
 contribution to the damping rate $T_1^{-1}$ is given by the simple formula
 \begin{equation}
T_1^{-1} = k^2 Q_T {\epsilon^2\over \omega_0^2} {L_qI_q^2 \over \hbar} \coth\left({\Omega_R \over 2T}\right).
\end{equation}
With $k^2Q_T = 0.1, \epsilon/\Delta = 0.1$ we find that the Rabi oscillations of the qubit are
terminated for the time $ T_1 = 60$ nsec at low temperature $T \ll \Omega_R$. The steady-state
polarization of the qubit is established for the shorter time $\Gamma_z^{-1} \simeq (1/2)T_1 \simeq
30$ nsec. Under the above-mentioned conditions it is possible to observe the Rabi oscillations of the
qubit resonantly coupled to the tank ($\Omega_R = \omega_T$) when, for example,  the qubit has the
tunneling frequency  $\Delta = 10$ GHz, and the Rabi frequency $\Omega_R$ is about 1 GHz. The qubit undergoes many Rabi oscillations during the relaxation time, $\Omega_R T_1 \simeq 60 \gg 1,$ that justifies the application of the perturbation theory to the case of resonant coupling of the qubit to the LC circuit having the finite quality factor ($Q_T \sim 1000$). Otherwise, it is better to work out of the resonant conditions between the qubit and the tank.

\section{Decoherence suppression in a strongly driven flux qubit.}

Coherent Rabi oscillations in a flux qubit have been observed recently by Chiorescu {\it et al} \cite{Chior2003}. Relaxation and dephasing times, $\tau_{relax}$ and $\tau_{\varphi},$ of the undriven qubit have been measured in the process together with a decay time, $\tau_{Rabi}$, of the strongly driven system: $\tau_{relax} = 900 ns, \tau_{\varphi} = 20 ns, $ and $\tau_{Rabi} = 150 ns$. It is of great interest to analyze these data using the theoretical formulas obtained above. 

Comparing the Hamiltonian (2) with the Hamiltonian employed in Ref. \cite{Chior2003} we find for the Rabi frequency (7): $\Omega_R = (\Delta /2\omega_0) \epsilon_{mw}.$ 
Here $\omega_0 = \sqrt{\Delta^2 + \epsilon^2} $ is energy splitting of the qubit which is exactly equal to the circular frequency of the driving force, $2\pi f: \omega_0 = 2\pi f,$ and $\epsilon_{mw}$ is the bias-modulation amplitude linearly proportional to the amplitude of the microwave field \cite{footnote1}. 

We suppose here that decoherence, dephasing and relaxation of the qubit originate from bias fluctuations $Q(t)$. These fluctuations, which also induce fluctuations of the qubit energy splitting, are characterized by the spectral density $S(\omega)$ (14) and the susceptibility $\chi(\omega)$ (15). 
Without the external driving force the dephasing and relaxation rates, $T_{2,eq}^{-1}$ and $T_{1,eq}^{-1}$, respectively, are proportional to the spectral density $S(\omega)$ (1):
\begin{eqnarray}
T_{1,eq}^{-1} = {\Delta^2\over 2\omega_0^2}S(\omega_0) , \nonumber\\
T_{2,eq}^{-1} = {1\over 2}T_{1,eq}^{-1} + {\epsilon^2 \over 2\omega_0^2} S(\omega \simeq 0).
\end{eqnarray}
These time scales correspond to the equilibrium relaxation time $T_{1,eq} = \tau_{relax}$ and to the free-evolution dephasing time $T_{2,eq} = \tau_{\varphi}$ measured by Chiorescu {et al.}\cite{Chior2003}.  
The pronounced difference between them ($\tau_{relax}^{-1} = 1.11 \times 10^6 s^{-1}, \tau_{\varphi}^{-1} = 50 \times 10^6 s^{-1}$) points to the fact that the 
slow fluctuations of the bias give a dominant contribution  ($\sim S(\omega \simeq 0)$)
into dephasing of the flux qubit. Fortunately, the low-frequency noise does not contribute to the equilibrium relaxation rate $T_{1,eq}^{-1} $. However, it is not the case for the driven qubit. It follows from Eq.(25) that  the nonequilibrium decay rate $T_1^{-1}$, corresponding to the decay rate $\tau_{Rabi}^{-1}$ \cite{Chior2003}, contains the term proportional to the spectrum of environment fluctuations at the Rabi frequency $S(\Omega_R):$
\begin{equation} 
T_1^{-1} = {3\Delta^2 \over 8 \omega_0^2}  S(\omega_0)
+ {\epsilon^2\over 4 \omega_0^2} S(\Omega_R).
\end{equation}
Here we have taken into account that the Rabi frequency $\Omega_R$ is much less than the energy splitting $\omega_0: \Omega_R \ll \omega_0 $ (from Ref.\cite{Chior2003} we have $\Omega_R /2\pi \sim 100 MHz, \omega_0/2\pi \sim \Delta/2\pi \simeq 3.4 GHz $). 
The presence of low-frequency term in the decay rate (34) can explain the substantial difference between the equilibrium relaxation rate, $T_{1,eq}^{-1} = 1.11 \times 10^6 s^{-1} $, and the decay rate of the driven qubit, $T_1^{-1} = \tau_{Rabi}^{-1} = 6.66 \times 10^6 s^{-1}.$  This decay rate incorporates the high-frequency contribution 
$(3\Delta^2/8\omega_0^2)S(\omega_0) = (3/4)\tau_{relax}^{-1} = 0.83 \times 10^6 s^{-1}$ 
as well as a term $ (\varepsilon^2 /4 \omega_0^2) S(\Omega_R) = 5.83 \times 10^6 s^{-1}$ which is determined by the spectrum of noise fluctuations taken at the Rabi frequency.     

It should be emphasized that the contribution of low-frequency fluctuations into the dephasing rate of the undriven qubit $T_{2,eq}^{-1}$, which is proportional to the spectrum at almost zero frequency, $ S(\omega \simeq 0) $,  is much higher: $ (\epsilon^2 /2\omega_0^2) S(\omega \simeq 0) \simeq \tau_{\varphi}^{-1} - 0.5 \tau_{relax}^{-1} \simeq 49.4 \times 10^6 s^{-1}.$ From these data we can extract ratio between two values of the spectrum of environment fluctuations: 
\begin{equation}
{S(\omega \simeq 0)\over S(\Omega_R)} = {2 \tau_{\varphi}^{-1} - \tau_{relax}^{-1} \over 4\tau_{Rabi}^{-1} - 3\tau_{relax}^{-1} } \simeq 4.2. 
\end{equation}
These estimations suggest a pronounced frequency dispersion of the spectrum $S(\omega)$ in the range from almost zero frequencies up to the frequencies of order $\Omega_R/2\pi \simeq 100 MHz.$  A drastic decrease in the spectrum of environment fluctuations points to the fact that the effective correlation time $\tau_c$ of the environment, $\tau_c$, 
is more than the period of Rabi oscillations $(\sim 10 ns)$, $\tau_c > 10 ns$. 
However, the correlation time $\tau_c$ is probably less than 100 ns, as it follows from the results of spin-echo experiments carried out by Chiorescu {\it et al.} \cite{Chior2003}. It would be interesting to measure a dependence of the decay time $\tau_{Rabi}$ on the amplitude of the external microwave field in the whole range of Rabi frequencies to restore a frequency dependence of the spectrum $S(\omega)$. 
For the flat spectrum, when $ S(\Omega_R) = S(\omega \simeq 0),$ the decay rate would be significantly higher: $T_{1,flat}^{-1} = 25.3 \times 10^6 s^{-1}$ (compared to $\tau_{Rabi}^{-1} = 6.66 \times 10^6 s^{-1}$), so that for the decay time of the driven qubit we would have : $T_{1,flat} = \tau_{Rabi,flat} = 39.5 ns = \tau_{Rabi}/3.8$ (in a comparison with the experimentally measured value $\tau_{Rabi} = 150 ns$). 
The difference between  $\tau_{Rabi}$ and $\tau_{Rabi,flat}$ can be interpreted as a result of decoherence suppression (in 4 times) by the strong microwave field acting on the flux qubit. The reason for this suppression lies in the fact that the low-frequency noise is averaged over the fast Rabi oscillations. 

Finally, we have to mention that for the strongly driven qubit it is not clear how to define separately the dephasing and the relaxation times. Here again we have two time scales $T_1$ (25) and $\Gamma_z^{-1}$ (19) which are related according to Eq.(29). It follows from the experimental data \cite{Chior2003} that the level populations of the flux qubit are equalized for the time $T_1 = 150 ns$ (26), whereas establishing of steady-state polarization of the qubit in the rotating frame of reference (27) is determined by both time scales, $T_1$ and $\Gamma_z^{-1}$, with $\Gamma_z^{-1} \simeq 80 ns.$ 

\section{Conclusions}
We have considered a dissipative dynamics of the two-state system (the qubit) driven by the strong resonant field. The generalized Bloch equations (17) have
been derived for the case of a weak interaction between the dynamical system and the heat bath. We have shown that a decay of Rabi oscillations and an establishment of the qubit  polarization along the external oscillating field
are determined by two different damping rates $T_1^{-1}$ (25) and $\Gamma_z$ (19) which are related by the simple expression (29). These results are true for weak coupling between the qubit and the heat bath and are not restricted to the case of Gaussian statistics of the heat bath fluctuations. 
It is demonstrated also 
that the Rabi oscillations eliminate the divergency in the decoherence rate of the biased qubit caused by the $1/f$ noise. We have analyzed the necessary conditions for the observations of the Rabi oscillations in the phase qubit coupled to the LC-circuit having the resonant frequency which is equal to the Rabi frequency. It is emphasized that recent measurements of decoherence and relaxation times in a driven flux qubit can be regarded as an experimental evidence of decoherence (dephasing) suppression by the strong resonant field.

\begin{center}
{\bf Acknowledgement}
\end{center}
Valuable discussions with M.H.S. Amin, Ya. Greenberg, A. Maassen van den Brink, A. Shnirman, and A. Zagoskin are
gratefully acknowledged. I am especially thankful to Alexandre Zagoskin for critical reading of the manuscript.

\end{document}